\documentclass[conference]{IEEEtran}
\IEEEoverridecommandlockouts
\usepackage{cite}
\usepackage{amsmath,amssymb,amsfonts}
\usepackage{algorithmic}
\usepackage{graphicx}
\usepackage{textcomp}
\usepackage{xcolor}
\def\BibTeX{{\rm B\kern-.05em{\sc i\kern-.025em b}\kern-.08em
    T\kern-.1667em\lower.7ex\hbox{E}\kern-.125emX}}

\begin{document}

\title{On Comparison of Encoders for Attention based End to End Speech Recognition in Standalone and Rescoring Mode\\
}

\author{\IEEEauthorblockN{Raviraj Joshi}
\IEEEauthorblockA{
\textit{Flipkart Internet Pvt. Ltd.}\\
Bengaluru, India \\
raviraj.j@flipkart.com}
\and
\IEEEauthorblockN{Subodh Kumar}
\IEEEauthorblockA{
\textit{Flipkart Internet Pvt. Ltd.}\\
Bengaluru, India \\
subodh.kumar@flipkart.com}
}

\maketitle

\begin{abstract}
The streaming automatic speech recognition (ASR) models are more popular and suitable for voice-based applications. However, non-streaming models provide better performance as they look at the entire audio context. To leverage the benefits of the non-streaming model in streaming applications like voice search, it is commonly used in second pass re-scoring mode. The candidate hypothesis generated using steaming models is re-scored using a non-streaming model. 

In this work, we evaluate the non-streaming attention-based end-to-end ASR models on the Flipkart voice search task in both standalone and re-scoring modes. These models are based on Listen-Attend-Spell (LAS) encoder-decoder architecture. We experiment with different encoder variations based on LSTM, Transformer, and Conformer. We compare the latency requirements of these models along with their performance. Overall we show that the Transformer model offers acceptable WER with the lowest latency requirements. We report a relative WER improvement of around 16\% with the second pass LAS re-scoring with latency overhead under 5ms. We also highlight the importance of CNN front-end with Transformer architecture to achieve comparable word error rates (WER). Moreover, we observe that in the second pass re-scoring mode all the encoders provide similar benefits whereas the difference in performance is prominent in standalone text generation mode. 

\end{abstract}

\begin{IEEEkeywords}
automatic speech recognition, voice search, LAS, CTC, transformer, lstm, conformer, re-scoring
\end{IEEEkeywords}

\section{Introduction}

The end-to-end speech recognition models directly convert audio features or raw waveform into output text. These models provide easy training and inference setup as compared to traditional models with separate acoustic, pronunciation, and language components \cite{he2019streaming}. These end-to-end models are based on CTC, RNN-T, or LAS architectures \cite{graves2013speech, graves2012sequence, chan2015listen}. One of the important aspects influencing the choice of these models is the streaming capability. The streaming models transcribe the audio on the go whereas the non-streaming models wait for the entire audio before starting the transcription. 

The streaming models are desirable for consumer-facing applications like voice search \cite{rao2017exploring,li2020towards}. Immediate feedback in the form of interpreted text makes such applications more user-friendly. The non-streaming models can be utilized for applications like automated customer experience support. Although non-streaming models have limited applications it provides superior performance as compared to streaming models \cite{chan2016listen}. The streaming models make use of only past context to generate current text. The non-streaming models look at the entire audio context before producing text and hence provide better results as compared to limited context streaming models.

Recently, non-streaming models have been used in second pass re-scoring mode to rank the text hypothesis generated using the streaming models. The re-scoring happens once the entire audio is available. This combination is shown to improve the results of streaming models \cite{sainath2020streaming}. These second pass models are commonly based on Listen-Attend-Spell (LAS) architecture. The LSTM and Transformer based variants have been explored in literature. The LAS model attends to encoder output to score the target text. The successor to these re-scoring systems are the recently proposed deliberation networks \cite{hu2020deliberation}. These networks encode the candidate hypothesis using a separate text encoder and the decoder attends to both audio and text representations. This allows the network to utilize bidirectional context with both text and audio. However, this further complicates the architecture and hence is not considered in this work.

In this work, we describe the LAS-based non-streaming models and evaluate them on Flipkart Voice Search task \cite{joshi2021attention,joshi2022simple}. The models are evaluated in standalone text generation mode and second pass rescoring mode. The two-pass system uses hypotheses generated from the CTC-based streaming model. We present a comparative analysis of different encoder architectures based on LSTM, Transformer, and Conformer. These architectures have provided state-of-the art results over time and hence been considered in this work. We also explore a shared first-pass system encoder instead of an independent audio encoder. We study these variations in re-scoring mode and first-pass standalone mode and make some important observations. To the best of our knowledge this is the first work to perform such a comparison for LAS based ASR systems.
\section{Related Work}

The most basic form of second pass re-scoring involves the usage of language models. The language models (LM) trained on large text-only corpus have even been shown to improve the accuracy of end-to-end ASR systems \cite{chiu2018state}. The text-only corpus is easily available as compared to parallel audio-text pairs and can be leveraged to train language models. The language models are used during the decoding of text as well as to re-score the final hypothesis \cite{kannan2018analysis}. The importance of using more powerful BERT-based language models has been shown in \cite{shin2019effective}. However, the BERT-based re-scoring has not been popular due to latency requirements. A more usable approach incorporating discriminative loss during BERT training was proposed recently in \cite{xu2022rescorebert}.

A class of text correction models has been proposed to correct the generated ASR text \cite{hrinchuk2020correction,wang2020asr}. The idea is similar to having a second pass system however it takes a step further to correct the text instead of just re-scoring the candidate hypothesis.

On similar lines, a second pass ASR model can be used to re-score the first pass hypothesis. The full second pass ASR model performs better than language models as it also considers the full audio context. However, both forms of re-scoring provide complementary benefits and are often used together. A two-pass model using streaming RNN-T architecture in the first pass and a non-streaming LAS in the second pass was proposed in \cite{sainath2020streaming}. They showed that RNN-T + LAS architecture outperformed the conventional streaming models while providing acceptable latency. The LSTM based LAS decoder was replaced with a Transformer decoder in \cite{li2020parallel} to enable efficient utilization of resources. The deliberation models were used for second pass rescoring in \cite{hu2020deliberation}. The deliberation models attend to both audio representation and text representations and are shown to work better than LAS rescoring which attends to audio representations only. Further, the LSTM decoder in the deliberation network was replaced with a Transformer decoder in \cite{hu2021transformer}.

\begin{figure}[h]
  \centering
  \includegraphics[scale=0.5]{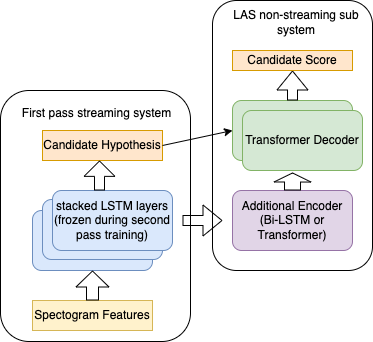}
  \caption{The two pass flow with shared encoder}
  \label{fig:shared_encoder}
\end{figure}

\begin{figure}[h]
  \centering
  \includegraphics[scale=0.5]{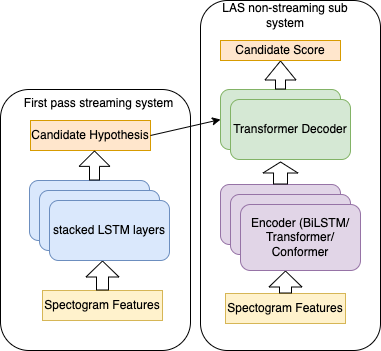}
  \caption{The two pass flow with independent encoder}
  \label{fig:independent_encoder}
\end{figure}

\section{Methodology}

We evaluate the non-streaming LAS models in standalone and re-scoring mode. The rescoring mode follows a two-step process. The first pass top 100 hypothesis generated using streaming CTC based system is re-scored using non-streaming LAS model. In the standalone mode, we do a beam search with beam size = 10. The process for rescoring mode is shown in Figure \ref{fig:shared_encoder} and Figure \ref{fig:independent_encoder}. The rescoring mode is evaluated in the shared encoder and independent encoder setup. The shared encoder setup utilizes the trained encoder of the first pass system with some extra layers. The encoder of the independent encoder system is different from the first pass system and hence trained from scratch. 

The LAS model follows an encoder-decoder architecture. The decoder attends to the encoder hidden representation and generates the output text sequentially. During re-scoring mode entire candidate hypothesis is processed in parallel to generate the score \cite{li2020parallel}. The log probability of the input token sequence is taken as the LAS score. 
The LAS score is combined with the other scores using the following formula:
\begin{align}
\begin{split}
    final\_score = \lambda_1 &* ctc\_score + \lambda_2 * las\_score \\
    &+ \lambda_3 * lm\_score + \lambda_4 * len
\end{split}
\end{align}
The values of $\lambda_1$, $\lambda_2$, $\lambda_3$, and $\lambda_4$ are fixed using grid search on the held-out validation data. The $ctc\_score$ indicates the score from first-pass model, $lm\_score$ is the language model (LM) score from a domain LM and $len$ is the number of subword tokens.

The decoder architecture is fixed to a 2 layer transformer. We explore different encoders for the LAS model based on LSTM, Transformer, and Conformer in both modes. We also explore the integration of the CTC model encoder in the LAS setup. This setup is similar to the method outlined in \cite{hu2020deliberation}. The LSTM encoder used in the first pass model is frozen and used as the LAS encoder.

\begin{table*}
  \centering
  \caption{Word Error Rate(WER) and encoder latency(ms) for different LAS encoder variations. The first pass model is a CTC-based system (encoder only), all others are LAS variants (encoder-decoder). The last two rows correspond to the shared encoder setup where the pre-trained encoder of the first pass system is used in the LAS setup. The weights of the pre-trained encoder are frozen during training. Moreover, they do not contribute to the trainable parameters as well. The decoder of all the LAS variants is a 2-layer Transformer.}
  \begin{tabular}{cccccl}
    \hline
    \textbf{Model} & \textbf{standalone mode} & \textbf{re-scoring mode} & \textbf{avg latency (ms)} & \textbf{trainable params (M)} \\
    \hline
    First Pass (CTC LSTM) Model & 9.11 & - & - & - \\ \hline
    Bi-LSTM Encoder & \textbf{7.65} & 7.47 & 22.31 & 56.5\\ \hline
    Transformer Encoder & 7.89 & 7.55 & 10.18 & 59.8\\ \hline
    VGG-CNN + Transformer Encoder& 7.69 & 7.41 & 12.22 & 61.3\\ \hline
    Conformer Encoder & 7.91 & \textbf{7.39} & 30.57 & 88.3 \\ \hline
    (CTC LSTM Encoder +) 2 BiLSTM & 8.19 & 7.73 & 11.79 & 33.1\\ \hline
    (CTC LSTM Encoder +) 4 Transformer & 8.45 & 7.65 & \textbf{4.25} & 34.9 \\
\hline
\end{tabular}
  \label{tab:all_res}
\end{table*}

\subsection{Model Architecture}
All the models outlined below use the same pre-processing steps \cite{joshi2021attention}. The standard log-mel features are input to the model. The 80-dimensional features are computed using a window size of 20ms and an overlap of 10 ms. The log-mel spectrogram is further subjected to time and frequency masking using spec-augment \cite{park2019specaugment}. The three consecutive time-steps are stacked and 240 dim features are input to the models. The models with CNN frontend do not use feature stacking. We use sub-words as the output units for all the models with a vocabulary size of 5000 \cite{kudo2018subword}. We train a custom unigram Google sentence piece tokenizer \cite{kudo2018sentencepiece} on the ASR text data. In order to ensure a fair comparison of models, we use standard architectures with minimal variations. A validation set is used to pick the best hyper parameters individually for all models.
\begin{itemize}
    \item \textbf{Bi-LSTM Encoder}: It uses 5 Bi-LSTM layers (512 units) with pyramidal compression \cite{chan2016listen} after every second layer. The pyramidal structure gives us 4x compression in the number of time steps. The scheduled sampling rate of 0.3 was used during training. Label smoothing is also used for all the models used in this work. 
    \item \textbf{Transformer Encoder}: It uses standard transformer self-attention blocks with an internal dimension of 512 units and the feed-forward dimension of 2048 units. It uses scaled dot product attention with 4 attention heads having 1024 units each. A total of 10 transformer blocks are stacked on top of each other. 
    \item \textbf{VGG-CNN + Transformer Encoder}: This model adds a VGG-CNN front end before the transformer encoder blocks as described in \cite{hori2017advances}. The VGG-CNN block has two CNN layers with a kernel size of 3x3 and a filter size of 64 followed by a max-pooling layer of size 2x2. This is again followed by two CNN layers of kernel size 3x3 and filter size 128 and a 2x2 max-pooling layer. The output of each time step is converted into a 512 dim vector using a dense layer before passing it to transformer layers. Except for the initial VGG-CNN block, the entire model is the same as the Transformer Encoder model described above. Input features are not stacked for this configuration. 
    \item \textbf{Conformer Encoder}: It uses stacked standard Conformer blocks as proposed in \cite{gulati2020conformer}. However, we make use of 10 Conformer layers with encoder dimension as 512 in order to make the configuration comparable with the Transformer encoder.
    \item \textbf{CTC-LSTM Frozen Encoder}: It consists of 12 stacked uni-directional LSTM layers. All the layers use a hidden dimension of 700 units. The layers are trained during the first pass training using the CTC objective. This model is frozen and used as the encoder of the LAS model. On top of the frozen encoder, we also add an additional small encoder. The additional encoder either uses 2 Bi-LSTM layers or 4 Transformer layers.  
    \item \textbf{Transformer Decoder}: The same transformer decoder is used with all the encoder forms. It uses 2 standard transformer decoder blocks with masked self-attention and encoder-decoder attention. The model dimensions and attention structure are the same as the transformer encoder. The decoder is trained to generate the sub-word units using teacher forcing \cite{bahdanau2015neural}. The cross-entropy loss function is used for training. 
\end{itemize}

\subsection{Dataset Details}
The primary data source used in this work is the Flipkart voice search (VS) data. The audio data is acquired through the Flipkart mobile application and manually transcribed by trained annotators. The dataset consists of around 6 million sentences with equal distribution of Hindi and English VS queries. We approximately have 8000 hours of VS training data. Along with the VS data, we have 4 million general domain crowdsourced audio-text pairs. This amounts to approx 6500 hours of general domain data. The validation set and test set consists of 10k and 22k samples from the VS domain. 

\section{Results}
  In this work, we report word error rates for different LAS models in standalone and rescoring mode. For each mode, we evaluate LSTM, Transformer, and Conformer-based encoders. Another efficient approach is to re-use the first pass encoder for second pass re-scoring. The results for standalone mode and re-scoring are shown in Table \ref{tab:all_res}. In the standalone mode, the Bi-LSTM pyramidal encoder and Transformer encoder with VGG frond-end perform the best. This shows the importance of the VGG frontend with the Transformers model. Although Conformer has been shown to produce state-of-the-art results on Libri Speech data, the numbers are slightly on the lower side when used as a LAS encoder on our VS dataset. The Conformer model evaluated in this work uses 10 encoder layers as opposed to 17 used in the original work. This allows us to train the model on commodity GPU systems. Finally, sharing the encoder with the first pass streaming system reduces the performance even further. We add BiLSTM and Transformer based additional small encoders on top of the frozen CTC encoder. The BiLSTM based system performs better than the Transformers counterpart in standalone mode.
 
 In re-scoring mode, the Conformer encoder performs the best but has the highest latency requirements. In general full encoders perform better than the shared encoder setup. However, the difference in WERs for different encoders is much less in re-scoring mode. Therefore the choice of encoder architecture is not much relevant and all the encoders perform more or less the same. The latency requirements for these encoders are also listed in Table \ref{tab:all_res}. The latency was computed on a Intel Xeon Skylake system with Nvidia A100 GPU. The transformers provide significant benefit in latency as compared to LSTM based models. Overall we report a relative WER improvement of 16-17\% over the first pass streaming system. The latency requirements of the shared encoder setup are the lowest as we only consider computations in the additional encoder part. The shared encoder setup provides comparable WERs with minimal latency requirements.

\section{Conclusion}
In conclusion, we evaluate different variations of non-streaming LAS models on the Flipkart Voice Search Task. The end-to-end speech recognition models were evaluated in standalone and re-scoring mode. We compare different encoders based on Bi-LSTM, Transformer, Conformer, and shared uni-LSTM. The Transformer model with the VGG frontend is shown to provide the best accuracy with reasonable latency requirements. In the re-scoring mode, different encoders provide very competitive performance as the hypothesis is generated by the first pass model. The lowest latency approach with an overhead of 4.25ms provides a relative WER improvement of 16\% over the first pass system.

\bibliographystyle{IEEEtran}
\bibliography{main}

\begin{thebibliography}{10}
\providecommand{\url}[1]{#1}
\csname url@samestyle\endcsname
\providecommand{\newblock}{\relax}
\providecommand{\bibinfo}[2]{#2}
\providecommand{\BIBentrySTDinterwordspacing}{\spaceskip=0pt\relax}
\providecommand{\BIBentryALTinterwordstretchfactor}{4}
\providecommand{\BIBentryALTinterwordspacing}{\spaceskip=\fontdimen2\font plus
\BIBentryALTinterwordstretchfactor\fontdimen3\font minus
  \fontdimen4\font\relax}
\providecommand{\BIBforeignlanguage}[2]{{%
\expandafter\ifx\csname l@#1\endcsname\relax
\typeout{** WARNING: IEEEtran.bst: No hyphenation pattern has been}%
\typeout{** loaded for the language `#1'. Using the pattern for}%
\typeout{** the default language instead.}%
\else
\language=\csname l@#1\endcsname
\fi
#2}}
\providecommand{\BIBdecl}{\relax}
\BIBdecl

\bibitem{he2019streaming}
Y.~He, T.~N. Sainath, R.~Prabhavalkar, I.~McGraw, R.~Alvarez, D.~Zhao,
  D.~Rybach, A.~Kannan, Y.~Wu, R.~Pang \emph{et~al.}, ``Streaming end-to-end
  speech recognition for mobile devices,'' in \emph{ICASSP 2019-2019 IEEE
  International Conference on Acoustics, Speech and Signal Processing
  (ICASSP)}.\hskip 1em plus 0.5em minus 0.4em\relax IEEE, 2019, pp. 6381--6385.

\bibitem{graves2013speech}
A.~Graves, A.-r. Mohamed, and G.~Hinton, ``Speech recognition with deep
  recurrent neural networks,'' in \emph{2013 IEEE international conference on
  acoustics, speech and signal processing}.\hskip 1em plus 0.5em minus
  0.4em\relax Ieee, 2013, pp. 6645--6649.

\bibitem{graves2012sequence}
A.~Graves, ``Sequence transduction with recurrent neural networks,''
  \emph{arXiv preprint arXiv:1211.3711}, 2012.

\bibitem{chan2015listen}
W.~Chan, N.~Jaitly, Q.~V. Le, and O.~Vinyals, ``Listen, attend and spell,''
  \emph{arXiv preprint arXiv:1508.01211}, 2015.

\bibitem{rao2017exploring}
K.~Rao, H.~Sak, and R.~Prabhavalkar, ``Exploring architectures, data and units
  for streaming end-to-end speech recognition with rnn-transducer,'' in
  \emph{2017 IEEE Automatic Speech Recognition and Understanding Workshop
  (ASRU)}.\hskip 1em plus 0.5em minus 0.4em\relax IEEE, 2017, pp. 193--199.

\bibitem{li2020towards}
B.~Li, S.-y. Chang, T.~N. Sainath, R.~Pang, Y.~He, T.~Strohman, and Y.~Wu,
  ``Towards fast and accurate streaming end-to-end asr,'' in \emph{ICASSP
  2020-2020 IEEE International Conference on Acoustics, Speech and Signal
  Processing (ICASSP)}.\hskip 1em plus 0.5em minus 0.4em\relax IEEE, 2020, pp.
  6069--6073.

\bibitem{chan2016listen}
W.~Chan, N.~Jaitly, Q.~Le, and O.~Vinyals, ``Listen, attend and spell: A neural
  network for large vocabulary conversational speech recognition,'' in
  \emph{2016 IEEE international conference on acoustics, speech and signal
  processing (ICASSP)}.\hskip 1em plus 0.5em minus 0.4em\relax IEEE, 2016, pp.
  4960--4964.

\bibitem{sainath2020streaming}
T.~N. Sainath, Y.~He, B.~Li, A.~Narayanan, R.~Pang, A.~Bruguier, S.-y. Chang,
  W.~Li, R.~Alvarez, Z.~Chen \emph{et~al.}, ``A streaming on-device end-to-end
  model surpassing server-side conventional model quality and latency,'' in
  \emph{ICASSP 2020-2020 IEEE International Conference on Acoustics, Speech and
  Signal Processing (ICASSP)}.\hskip 1em plus 0.5em minus 0.4em\relax IEEE,
  2020, pp. 6059--6063.

\bibitem{hu2020deliberation}
K.~Hu, T.~N. Sainath, R.~Pang, and R.~Prabhavalkar, ``Deliberation model based
  two-pass end-to-end speech recognition,'' in \emph{ICASSP 2020-2020 IEEE
  International Conference on Acoustics, Speech and Signal Processing
  (ICASSP)}.\hskip 1em plus 0.5em minus 0.4em\relax IEEE, 2020, pp. 7799--7803.

\bibitem{joshi2021attention}
R.~Joshi and V.~Kannan, ``Attention based end to end speech recognition for
  voice search in hindi and english,'' in \emph{Forum for Information Retrieval
  Evaluation}, 2021, pp. 107--113.

\bibitem{joshi2022simple}
R.~Joshi and A.~Singh, ``A simple baseline for domain adaptation in end to end
  asr systems using synthetic data,'' in \emph{Proceedings of The Fifth
  Workshop on e-Commerce and NLP (ECNLP 5)}, 2022, pp. 244--249.

\bibitem{chiu2018state}
C.-C. Chiu, T.~N. Sainath, Y.~Wu, R.~Prabhavalkar, P.~Nguyen, Z.~Chen,
  A.~Kannan, R.~J. Weiss, K.~Rao, E.~Gonina \emph{et~al.}, ``State-of-the-art
  speech recognition with sequence-to-sequence models,'' in \emph{2018 IEEE
  International Conference on Acoustics, Speech and Signal Processing
  (ICASSP)}.\hskip 1em plus 0.5em minus 0.4em\relax IEEE, 2018, pp. 4774--4778.

\bibitem{kannan2018analysis}
A.~Kannan, Y.~Wu, P.~Nguyen, T.~N. Sainath, Z.~Chen, and R.~Prabhavalkar, ``An
  analysis of incorporating an external language model into a
  sequence-to-sequence model,'' in \emph{2018 IEEE International Conference on
  Acoustics, Speech and Signal Processing (ICASSP)}.\hskip 1em plus 0.5em minus
  0.4em\relax IEEE, 2018, pp. 1--5828.

\bibitem{shin2019effective}
J.~Shin, Y.~Lee, and K.~Jung, ``Effective sentence scoring method using bert
  for speech recognition,'' in \emph{Asian Conference on Machine
  Learning}.\hskip 1em plus 0.5em minus 0.4em\relax PMLR, 2019, pp. 1081--1093.

\bibitem{xu2022rescorebert}
L.~Xu, Y.~Gu, J.~Kolehmainen, H.~Khan, A.~Gandhe, A.~Rastrow, A.~Stolcke, and
  I.~Bulyko, ``Rescorebert: Discriminative speech recognition rescoring with
  bert,'' \emph{arXiv preprint arXiv:2202.01094}, 2022.

\bibitem{hrinchuk2020correction}
O.~Hrinchuk, M.~Popova, and B.~Ginsburg, ``Correction of automatic speech
  recognition with transformer sequence-to-sequence model,'' in \emph{ICASSP
  2020-2020 IEEE International Conference on Acoustics, Speech and Signal
  Processing (ICASSP)}.\hskip 1em plus 0.5em minus 0.4em\relax IEEE, 2020, pp.
  7074--7078.

\bibitem{wang2020asr}
H.~Wang, S.~Dong, Y.~Liu, J.~Logan, A.~K. Agrawal, and Y.~Liu, ``Asr error
  correction with augmented transformer for entity retrieval.'' in
  \emph{Interspeech}, 2020, pp. 1550--1554.

\bibitem{li2020parallel}
W.~Li, J.~Qin, C.-C. Chiu, R.~Pang, and Y.~He, ``Parallel rescoring with
  transformer for streaming on-device speech recognition,'' \emph{Proc.
  Interspeech 2020}, pp. 2122--2126, 2020.

\bibitem{hu2021transformer}
K.~Hu, R.~Pang, T.~N. Sainath, and T.~Strohman, ``Transformer based
  deliberation for two-pass speech recognition,'' in \emph{2021 IEEE Spoken
  Language Technology Workshop (SLT)}.\hskip 1em plus 0.5em minus 0.4em\relax
  IEEE, 2021, pp. 68--74.

\bibitem{park2019specaugment}
D.~S. Park, W.~Chan, Y.~Zhang, C.-C. Chiu, B.~Zoph, E.~D. Cubuk, and Q.~V. Le,
  ``Specaugment: A simple data augmentation method for automatic speech
  recognition,'' \emph{arXiv preprint arXiv:1904.08779}, 2019.

\bibitem{kudo2018subword}
T.~Kudo, ``Subword regularization: Improving neural network translation models
  with multiple subword candidates,'' in \emph{Proceedings of the 56th Annual
  Meeting of the Association for Computational Linguistics (Volume 1: Long
  Papers)}, 2018, pp. 66--75.

\bibitem{kudo2018sentencepiece}
T.~Kudo and J.~Richardson, ``Sentencepiece: A simple and language independent
  subword tokenizer and detokenizer for neural text processing,'' in
  \emph{Proceedings of the 2018 Conference on Empirical Methods in Natural
  Language Processing: System Demonstrations}, 2018, pp. 66--71.

\bibitem{hori2017advances}
T.~Hori, S.~Watanabe, Y.~Zhang, and W.~Chan, ``Advances in joint ctc-attention
  based end-to-end speech recognition with a deep cnn encoder and rnn-lm,''
  \emph{Proc. Interspeech 2017}, pp. 949--953, 2017.

\bibitem{gulati2020conformer}
A.~Gulati, J.~Qin, C.-C. Chiu, N.~Parmar, Y.~Zhang, J.~Yu, W.~Han, S.~Wang,
  Z.~Zhang, Y.~Wu \emph{et~al.}, ``Conformer: Convolution-augmented transformer
  for speech recognition,'' \emph{Proc. Interspeech 2020}, pp. 5036--5040,
  2020.

\bibitem{bahdanau2015neural}
D.~Bahdanau, K.~H. Cho, and Y.~Bengio, ``Neural machine translation by jointly
  learning to align and translate,'' in \emph{3rd International Conference on
  Learning Representations, ICLR 2015}, 2015.

\end{thebibliography}
\end{document}